\documentclass[]{aa}  

\usepackage{natbib}
\bibpunct{(}{)}{;}{a}{}{,}
\usepackage{graphicx}
\usepackage{revsymb}	
\usepackage{amsmath}	
\usepackage[usenames]{color}	
\usepackage{epstopdf}	
\usepackage{txfonts}
\usepackage{amssymb}
\usepackage{color}
\usepackage[justification=centering]{caption}
\usepackage{geometry} 
\usepackage[parfill]{parskip} 
\usepackage{amssymb}
\usepackage{slantsc}
\usepackage{array}
\usepackage{url}
\usepackage{amsfonts}
\usepackage[colorlinks=true,linkcolor=blue, urlcolor=blue, citecolor=blue]{hyperref}
\usepackage{sidecap}
\usepackage{graphicx}
\usepackage{xcolor}
\usepackage{nicefrac}
\usepackage{subfigure}
\usepackage{epsfig}
\colorlet{rouge}{red!70!darkgray}

\begin{document}
\title{In-depth analysis of solar models with high-metallicity abundances and updated opacity tables}
\author{G. Buldgen\inst{1,2} \and A. Noels\inst{2} \and R. Scuflaire\inst{2} \and A. M. Amarsi\inst{3} \and N. Grevesse \inst{2,4}  \and P. Eggenberger\inst{1} \and J. Colgan \inst{5}\and C. J. Fontes \inst{5} \and V. A. Baturin\inst{6} \and A. V. Oreshina\inst{6} \and S. V. Ayukov\inst{6} \and P. Hakel \inst{5}\and D. P. Kilcrease \inst{5}}
\institute{D\'epartement d'Astronomie, Universit\'e de Gen\`eve, Chemin Pegasi 51, CH-1290 Versoix, Switzerland \and STAR Institute, Université de Liège, Liège, Belgium \and Theoretical Astrophysics, Department of Physics and Astronomy, Uppsala University, Box 516, 751 20 Uppsala, Sweden \and Centre Spatial de Liège, Université de Liège, Angleur-Liège, Belgium \and Los Alamos National Laboratory, Los Alamos, NM 87545, USA \and Sternberg Astronomical Institute, Lomonosov Moscow State University, 119234,Moscow, Russia}
\date{October, 2023}
\abstract{As a result of the high quality constraints available for the Sun, we are able to carry out detailed combined analyses using neutrino, spectroscopic and helioseismic observations. Such studies lay the ground for future improvements of key physical components of solar and stellar models, as ingredients such as the equation of state, the radiative opacities or the prescriptions for macroscopic transport processes of chemicals are then used to study other stars in the Universe.}
{We aim at studying the existing degeneracies in solar models using the recent high-metallicity spectroscopic abundances by comparing them to helioseismic and neutrino data and discuss how their properties are impacted by changes in the micro and macro physical ingredients.}
{We carry out a detailed study of solar models computed with a high-metallicity composition from the literature based on averaged-3D models that has been claimed to resolve the solar modelling problem. We compare these models to helioseismic and neutrino constraints.}
{The properties of the solar models are significantly affected by the use of the recent OPLIB opacity tables and the inclusion of macroscopic transport. The properties of the standard solar models computed using the OPAL opacities are similar to those using the OP opacities. We show that a modification of the temperature gradient just below the base of the convective zone is required to erase the discrepancies in solar models, particularly in the presence of macroscopic mixing. This can be simulated by a localized increase of opacity of a few percent.}{We conclude that the existing degeneracies and issues in solar modelling are not erased by using an increase in the solar metallicity in contradiction to what has been suggested in recent literature.  Therefore, standard solar models cannot be used as an argument for a high metallicity composition.  While further work is required to improve solar models, we note that direct helioseismic inversions indicate a low metallicity in the convective envelope, in agreement with spectroscopic analyses based on full 3D models.}
\maketitle
\section{Introduction}

During the last 30 years, the solar metallicity, Z, has oscillated from a high value to a low value, to return again to a high value in a recently published paper. The early works by \citet{GrevNoels} (hereafter GN93) and \citet{GS1998} (hereafter GS98) led to values of Z/X=0.0244 and Z/X=0.0231 respectively. These results have been obtained from analysing spectra taken at the center of the solar disc using 1D LTE photospheric models. More recently, new analyses of the same solar spectra but using new atomic data and improved 3D NLTE models by \citet{AGSS09} (hereafter AGSS09) and \citet{Asplund2021} (hereafter AAG21) and \citet{Amarsi2021} derived much lower metallicities, Z/X=0.0181 and Z/X=0.0187 respectively. Very recently, \citet{Magg2022}(hereafter MB22) proposed a revision of the solar abundances leading to a metallicity of Z/X=0.0226, back to the high values of the 1990s and at 4.5 $\sigma$ with AAG21. This result is based on an analysis of the solar disc-integrated flux spectrum using a spatial and temporal average of a 3D RHD model, thus a 1D model called $<3\rm{D}>$. While further comparisons are required to fully understand the origin of the differences between the study of MB22 and those of AGSS09 and AAG21, it is however interesting to briefly compare 3D and $<3\rm{D}>$ models. The differences between these two types of models are now well known: the 3D model by far outperforms the $<3\rm{D}>$ model, as is clearly observed when applied to the analysis of disc-integrated flux spectra as shown in Fig. 7 of \citet{Amarsi2018} for the case of oxygen. The upwards revision of the metallicity by MB22 has rejuvenated the debate on the so-called ``solar problem”. As expected, their high metallicity value improves the situation with neutrino measurements and some helioseismic constraints. However, they only computed one set of standard solar models to draw these conclusions, leaving a detailed analysis of the solar models to be performed later, arguing that the remaining discrepancies could be explained by remaining limitations of the stellar models and pointing to the work of \citet{Buldgen2019} that was carried out using AGSS09 as well as  abundances using the neon revision of \citet{Young2018}, thus compatible with AAG21. 
 
Recently, \citet{Buldgen2023} criticized the claims of MB22 regarding the needs for revision of the physics of solar models. They showed that the agreement found by MB22 was due to a combination of physical ingredients and not solely to the abundances. They discussed the apparearance of various issues once macroscopic transport of chemicals was included to reproduce the lithium depletion in the Sun, independently of the parametrization used for the transport coefficient. They also showed that an accurate determination of the solar beryllium abundance was required to fully characterize macroscopic transport at the base of the convective zone (BCZ). They mentioned that MB22 did not consider recent helioseismic determinations of the chemical composition of the solar envelope \citep{Vorontsov13, BuldgenZ}, which were further improved in precision \citep{Buldgen2023Z}, providing in this last study an average over multiple reference models and datasets of Z=0.0138. These independent approaches to determine the solar chemical composition confirm the low metallicity value of AAG21.

However, \citet{Buldgen2023} did not consider the impact of varying the reference opacity tables and combined helioseismic inversions, not discussing the question of the remaining ``limitations'' mentioned in the conclusions of MB22. In this study, we use standard solar models (SSMs) and non-standard models including macroscopic mixing of chemicals using both the OPAL and OPLIB opacities and carry out a detailed investigations of calibrated models computed using the MB22 abundances, in a similar fashion to \citet{Buldgen2019} who did such a detailed analysis using the AGSS09 abundances as well as the neon revision of \citet{Young2018}, which later was confirmed by AAG21. We thus significantly extend the set of solar models computed with the MB22 abundances and discuss our findings regarding global parameters such as the position of the BCZ, the helium mass fraction in the convective zone, neutrino fluxes and combined helioseismic inversion of the squared adiabatic sound speed, the entropy proxy, and the Ledoux discriminant as well as the frequency separation ratios of low-degree modes.
We implemented a new diffusion coefficient to study macroscopic mixing below the convective envelope, derived from the asymptotic behaviour of the combined shear instability and magnetic Tayler instability in rotating solar models instead of the usual power law in density \citep[see e.g.][]{Proffitt1991,Richard1996,JCD2018,Buldgen2023}. This approach is linked to the solid-body rotation of the solar radiative interior \citep{bro89,Thompson1996,SchouRota}. As the inclusion of macroscopic transport reduces the extent of the solar convective zone, we also investigate the behaviour of solar models under the effects of adiabatic overshooting and localized increase of opacities that recover the helioseismic position of the base of the convective zone.

By combining all constraints available for the Sun, we carry out a detailed analysis of solar models using the MB22 abundances. We discuss the actual implications of the ``residual limitations'' of SSMs computed with revised high-metallicity solar abundances and disentangle the various contributors to these discrepancies in a similar fashion to \citet{Buldgen2019}, taking into account macroscopic transport, localized opacity modifications and overshooting at the base of the convective envelope. Our aim is here to further demonstrate the need for improvement of the physics of solar models and that, even if the MB22 abundances are taken at face value, our conclusions remain unchanged and detailed helioseismic analyses of solar models built using these abundances combined with various opacity tables and a new formalism for macroscopic transport reinforce such needs rather than `` alleviate'' them. 

\section{Solar models}\label{Sec:Models}

We computed solar models using the Liège stellar evolution code \citep{ScuflaireCles} using various physical ingredients as in \citet{Buldgen2019}. We used the recently suggested high-metallicity solar abundances (MB22) based on $<3\rm{D}>$ models, the latest version (v7\footnote{\url{http://crydee.sai.msu.ru/SAHA-S_EOS/}}) of the SAHA-S equation of state \citep{Gryaznov2006,Gryaznov2013}. We refer the reader to \citet{Buldgen2019} and references therein for similar comparisons between high and low-metallicity solar models, here we discuss only MB22 solar models. From \citet{Buldgen2019} and from previous references, it appears that the two main ingredients affecting the properties of solar models are the transport of chemical elements and the opacity tables. In this study, we analyze in detail for the first time the implication of the abundance revision by MB22 for various opacity tables available in the literature. We thus compute the first MB22 standard solar models using the OPAL \citep{OPAL} and OPLIB \citep{Colgan} opacity tables computed for this specific mixture, as well as models including macroscopic transport of chemicals reproducing the combined effects of hydrodynamic and magnetic instabilities due to the presence of rotation in the solar radiative zone (models denoted $\rm{D_{R}}$). We follow the work of \citet{Eggenberger2022}, who computed solar models both reproducing the lithium depletion and the internal rotation profile of the solar radiative zone. To do so, we use an asymptotic description of the transport coefficient of chemicals under the combined effects of meridional circulation, shear instability and the magnetic Tayler instability \citep{spr02}. 

As the Sun is a slow rotator, the dominant transport mechanism of chemicals due to rotation is the shear instability in the radiative layers of solar models resulting from the presence of significant radial differential rotation \citep{Zahn1992}. When including the Tayler magnetic instability\citep{spr02}, the radial differential rotation is regulated via an efficient transport of angular momentum. However, a critical radial gradient of rotation is required for the instability to operate and chemical gradients have an inhibiting effect on the apparition of this process. Therefore, the magnetic Tayler instability acts as an intermittent very efficient angular momentum transport that reduces the efficiency of the transport of chemicals by shear. One can thus estimate an asymptotic diffusion coefficient for the chemicals that is essentially the transport by shear, where the rotation gradient is the critical value at which the magnetic Tayler instability operates (since a larger radial rotation gradient would be quickly damped by the magnetic Tayler instability back to the critical value). 

Following this reasoning, we use the equation for the critical radial shear for the instability to operate
\begin{align}
\bigg| \frac{\rm{d} \ln \Omega}{\rm{d} \ln \rm{r}}\bigg| \geqslant \left(\frac{\rm{N}_{\mu}}{\Omega}\right)^{7/4}\left( \frac{\eta}{\rm{r^{2}N}_{\mu}}\right)^{1/4},
\end{align}
with $\Omega$, the angular rotation velocity, assumed constant and fixed to the helioseismic value of the solar radiative zone, $\rm{N}_{\mu}$ the chemical contribution to the Brunt-V\"ais\"al\"a frequency and $\eta$ the magnetic diffusivity, and combine it with the vertical diffusion coefficient of the shear instability of \citet{Talon1997}, $\rm{D_{X}}$, that takes consistently the effects of chemical composition gradients into account
 
\begin{align}
\rm{D_{X}}\approx \frac{2 Ri_{c}(dU/dz)^{2}}{\rm{N}^{2}_{\rm{T}}/(\rm{K+D_{h}})+N^{2}_{\mu}/\rm{D_{h}}}, \label{eq:ShearTalon}
\end{align}

with $\rm{D_{h}}$ the horizontal turbulence coefficient, $\rm{Ri_{c}}$ the critical Richardson number, $\rm{dU/dz}=\rm{r} \sin\theta (\rm{d}\Omega/\rm{dr})$, the vertical shear rate, K the thermal diffusivity and $\rm{N_{T}}$ the thermal contribution to the Brunt-V\"ais\"al\"a frequency. The following expression for the macroscopic transport of chemicals is obtained after averaging over latitude
\begin{align}
    \rm{D_{R}}=\rm{D_{h}f(r)}\Omega^{-3/2} \left(\frac{\eta \vert \rm{N}^{2}_{\mu} \vert}{\rm{r}^{2}}\right)^{1/2}, \label{Eq:DiffCoeff}
\end{align}
with $\rm{f(r)}$ a parametric function that is used to mimic the overall complex behaviour of the coefficient when the full transport of both angular momentum and chemicals is computed. The behaviour is, expectedly, very similar to the recalibrated density power-law used in \citet{Eggenberger2022}. We mention however that such an approach would need to be recalibrated in light of the incompatibility of the magnetic Tayler instability with the observations at later evolutionary stages \citep[see e.g.][]{Deheuvels2014,Deheuvels2015,Gehan2018}, particularly with the very young subgiants of \citet{Deheuvels2020}. It would however not impact the conclusions of our study as they are similar to those of \citet{Buldgen2023}

We compute eight models in total, as we also investigate the impact of replacing the position of the base of the convective zone at the helioseismically inferred value of $0.713\pm 0.001$ R$_{\odot}$ \citep{JCD1991,Basu97BCZ} using either adiabatic convective penetration (denoted ``Ov'') or a localized increase of opacity (denoted ``OPAC''). We investigate whether one of these solutions is favoured over the other in the context of helioseismic inversions of the solar structure. 

\begin{figure}
	\centering
		\includegraphics[width=9cm]{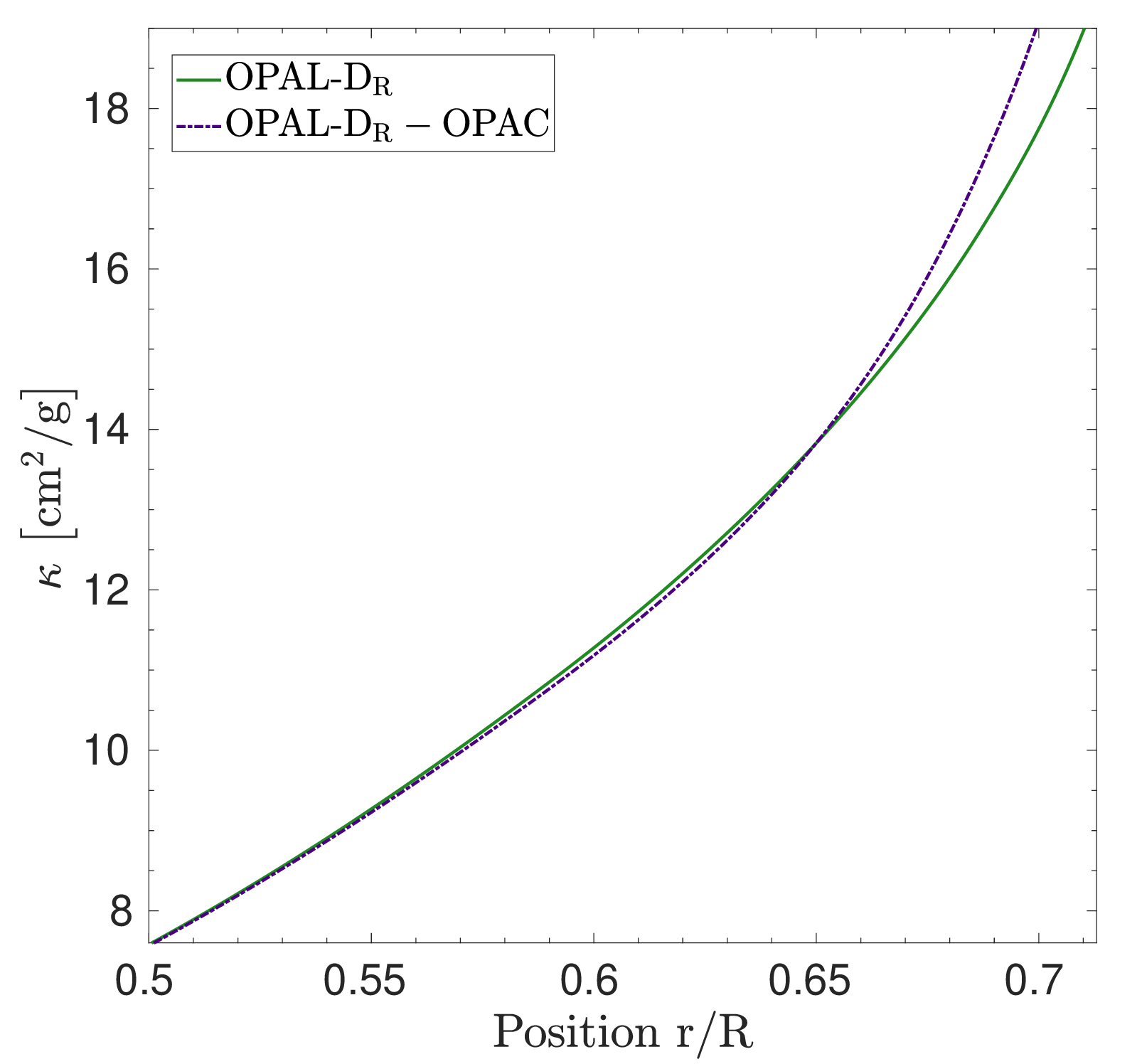}
	\caption{Opacity profile as a function of the normalized radius for Model OPAL $D_{R}$ (green) and Model OPAL $D_{R}$+OP (blue).}
		\label{Fig:Kappa}
\end{figure}

The increase in opacity is parametrized as follows
\begin{align}
\kappa=\kappa_{0}(1+\delta \kappa), \label{eq:OpalParam}
\end{align}
with $\kappa_{0}$ the reference opacity of the model (e.g. either OPAL or OPLIB) and $\delta \kappa$ the Gaussian opacity modification parametrized with temperature
\begin{align}
\delta \kappa = A\exp \left(-150 (\log T - 6.33)^{2} \right), \label{eq:OpalDelta}
\end{align}
with $A$ the amplitude of the modification and T the local temperature. The temperature used is close to that of the \citet{Bailey} experiment, with an extension sufficient to affect the radiative layers that are at slitghtly hotter temperatures in our models. The parametrization is kept constant throughout the evolution of the solar model and leads to a modification of the opacity profile at the BCZ illustrated in Fig. \ref{Fig:Kappa} for the OPAL model. Despite peaking in the convective zone (the BCZ is located at $\log T = 6.34$), the modifications in the radiative layers just below the BCZ are still substantial and lead to significant differences in helioseismic inference results.

\subsection{Global parameters}\label{Sec:Parameters}

We start by taking a look at the relevant global parameters describing solar models, namely the radial coordinate position of the base of the convective envelope, the mass coordinate at the position of the base of the convective envelope, the helium mass fraction in the convective envelope and the photospheric lithium abundance. The values of these various parameters for each model are provided in Table \ref{tabModels}.

\begin{table*}[h]
\caption{Global parameters of the solar evolutionary models}
\label{tabModels}
  \centering
\begin{tabular}{r | c | c | c | c }
\hline \hline
\textbf{Name}&\textbf{$\left(r/R\right)_{\rm{BCZ}}$}&\textbf{$\left( m/M \right)_{\rm{CZ}}$}&\textbf{$\mathit{Y}_{\rm{CZ}}$}&\textbf{$\mathrm{A}\left( Li\right)$ $\left( dex \right)$} \\ \hline
Model OPAL SSM&$0.7173$&$0.9770$& $0.2460$ & $2.536$\\
Model OPAL $\rm{D_{R}}$&$0.7210$&$0.9779$& $0.2545$ & $0.954$\\ 
Model OPAL $\rm{D_{R}+Ov}$&$0.7133$&$0.9777$& $0.2535$ & $0.915$\\
Model OPAL $\rm{D_{R}+OPAC}$&$0.7136$&$0.9762$& $0.2546$ & $0.918$\\
Model OPLIB SSM&$0.7142$&$0.9761$& $0.2404$ & $2.611$\\
Model OPLIB $\rm{D_{R}}$&$0.7185$&$0.9769$& $0.2484$ & $0.991$\\ 
Model OPLIB $\rm{D_{R}+Ov}$&$0.7133$&$0.9768$& $0.2479$ & $0.991$\\
Model OPLIB $\rm{D_{R}+OPAC}$&$0.7132$&$0.9757$& $0.2485$ & $0.982$\\
\hline
\end{tabular}
\end{table*}

A first conclusion drawn from Table \ref{tabModels} is that the results of \citet{Buldgen2023} regarding the helium mass fraction in the convective envelope $\mathit{Y}_{\rm{CZ}}$ hold for the OPAL and OPLIB opacities.  The OPAL tables were the reference opacity tables for the SSMs of the 1990s \citep[e.g.][]{JCD1996}, they have been replaced by the OP opacities \citep{Badnell2005} in recent SSMs \citep{Vinyoles2017}. The recent OPLIB opacities are the latest generation of Los Alamos opacities, investigated in \citet{Colgan2016} and \citet{BuldgenS}.

While the OPAL SSM is in excellent agreement in $\mathit{Y}_{\rm{CZ}}$ with respect to the helioseismic measurement of $\mathit{Y}_{\rm{CZ},\odot}=0.2485\pm0.0035$ \citep{BasuYSun}, the models including the effects of macroscopic mixing show a too high value with respect to that inferred from helioseismology. We also note that a more recent determination by \citet{Vorontsov13} using modern equations of state shows a slightly larger interval of values, favouring higher helium mass fraction values around $0.25$ in the CZ. This issue is still present in models for which the position of the base of the convective envelope is replaced at the helioseismic value ($0.713\pm0.001$R$_{\odot}$) using either overshooting or an opacity increase. The inclusion of macroscopic mixing is required to reproduce the lithium photospheric abundance, $\rm{A(Li)}=0.96\pm0.05$ dex \citep{Wang2021}, but reduces the size of the solar convective envelope and thus destroys the existing agreement of high metallicity models with helioseismology, as shown in Table \ref{tabModels}. To restore this agreement, either an adiabatic convective penetration of $0.088 H_{P}$ is applied at the BCZ (with $H_{P}$ the local pressure scale height) or an increase of opacity of $11\%$  at the BCZ is applied (namely $A=0.12$), following Equations \ref{eq:OpalParam} and \ref{eq:OpalDelta}.

The OPLIB models show a similar behaviour. However, due to the intrinsically lower values of the OPLIB opacities compared to the OPAL ones at high temperature, the $\mathit{Y}_{\rm{CZ}}$ values of the models are shifted by about $0.005$. This means that the OPLIB models including macroscopic transport provide an overall better agreement than the OPAL ones, particularly since they lead naturally to a deeper position of the base of the convective envelope. As discussed in \citet{Buldgen2023}, the BCZ position is also significantly affected by the details of the formalism used in the computation of microscopic diffusion of chemicals. From this work, as well as from \citep{Buldgen2019}, we see that the effects of the screening coefficients of \citet{Paquette} is to push the BCZ position up by about $0.003$. 

Therefore, if the original implementation of \citet{Thoul} was used as in MB22, a SSM using the OPLIB opacities would have slightly deeper position of the BCZ, close to be in disagreement with the helioseismic value. This illustrates nicely the degeneracy that exists in the solar models and the various parameters that can lead to agreeing (or not) with the extremely precise constraints for the Sun. Nevertheless, OPLIB models including macroscopic transport still need some increase of opacity or adiabatic convective penetration to replace the BCZ position at the helioseismic value. In this case the convective penetration is only of $0.061 H_{P}$ and the increase in opacity is of $8.0\%$ at the BCZ (or $A=0.085$ in Eq. \ref{eq:OpalDelta}).

\subsection{Neutrino fluxes}\label{Sec:Neut}

The second relevant constraints to investigate when computing solar models are the neutrino fluxes. The results for our models are summarized in Table \ref{tabNeutrinos}, where we illustrate the pp, Be, B and CNO neutrino fluxes, denoted respectively $\phi(\rm{pp})$, $\phi(\rm{B})$, $\phi(\rm{Be})$, $\phi(\rm{CNO})$. We compare these results to the values provided in \citet{Borexino2018} and \citet{OrebiGann2021}.  

\begin{table*}[h]
\caption{Neutrino fluxes of the evolutionary models}
\label{tabNeutrinos}
  \centering
\begin{tabular}{r | c | c | c | c }
\hline \hline
\textbf{Name}& $\phi(\rm{pp})$ $\left[ \times 10^{10}/\rm{cm}^{2}/\rm{s}\right]$& $\phi(\rm{Be})$ $\left[ \times 10^{9}/\rm{cm}^{2}/\rm{s}\right]$&$\phi(\rm{B})$ $\left[ \times 10^{6}/\rm{cm}^{2}/\rm{s}\right]$&$\phi(\rm{CNO})$ $\left[ \times 10^{8}/\rm{cm}^{2}/\rm{s}\right]$\\ \hline
Model OPAL Std& $5.94$ & $4.95$ & $5.53$ &$6.21$\\
Model OPAL $\rm{D_{R}}$& $5.97$ & $4.77$ & $5.12$ & $5.58$\\ 
Model OPAL $\rm{D_{R}+Ov}$& $5.97$ & $4.78$ & $5.15$ & $5.61$\\
Model OPAL $\rm{D_{R}+OPAC}$& $5.97$ & $4.77$ & $5.13$ & $5.57$\\
Model OPLIB Std& $5.98$ & $4.61$ & $4.58$ & $5.45$\\
Model OPLIB $\rm{D_{R}}$& $6.01$ & $4.44$ & $4.24$ & $4.92$\\
Model OPLIB $\rm{D_{R}+Ov}$ & $6.01$ & $4.45$ & $4.26$ & $4.94$ \\
Model OPLIB $\rm{D_{R}+OPAC}$ & $6.01$ & $4.45$ & $4.25$ & $4.92$ \\
O-G21$^{1}$ & $5.97^{+0.0037}_{-0.0033}$ & $4.80^{+0.24}_{-0.22}$ & $5.16^{+0.13}_{-0.09}$ & $-$\\
Borexino$^{2}$ & $6.1^{+0.6}_{-0.7}$ & $4.99^{+0.13}_{-0.14}$ & $5.68^{+0.39}_{-0.41}$ & $6.6^{+2.0}_{-0.9}$\\
\hline
\end{tabular}

\small{\textit{Note:} $^{1}$ \citet{OrebiGann2021}, $^{2}$ \citet{Borexino2018}, \citet{Borexino2020}, \citet{Appel2022}}
\end{table*}

The first point we confirm is that high metallicity SSMs computed with the OPAL opacities agree quite well with the Borexino fluxes, including the recent CNO fluxes of \citet{Appel2022}. As shown in Table \ref{tabNeutrinos}, we note however that the analysis of \citet{OrebiGann2021} provides a much lower $\phi_{\rm{B}}$ value than Borexino, leading to a significant disagreement with the high metallicity SSMs. As in \citet{Buldgen2023}, the inclusion of macroscopic transport leads to significant disagreement with the Borexino $\phi_{\rm{B}}$ value often used to favour high CNO abundances in the Sun \citep{Bahcall2005,Serenelli2013,Serenelli2016,Borexino2018}. In addition, the value of $\phi_{\rm{CNO}}$ is now also much lower, disagreeing at $1\sigma$ with the measurements. 

The issue is more tedious for the models computed with the OPLIB opacity tables. The SSM is already in disagreement at $1\sigma$ for the $\phi_{\rm{CNO}}$ measurements as well as with the Borexino measurements of $\phi_{\rm{B}}$ and $\phi_{\rm{Be}}$. Adding macroscopic mixing just leads to further increasing the disagreements, leading to questions about the properties of the solar core in the OPLIB models. The key parameter here is the lower opacity at high temperatures, that leads to a higher initial hydrogen abundance to reproduce the solar luminosity at the solar age. Therefore, for a given solar metallicity, the helium abundance and central temperature is lower, leading to a disagreement in neutrino fluxes and a lower helium mass fraction in the CZ. This effect on the neutrino fluxes is further increased by the inclusion of the effects of macroscopic mixing that has the tendency to push the calibration procedure towards even higher initial hydrogen abundances. 

\section{Helioseismic constraints}\label{Sec:Inversions}

To fully investigate the helioseismic properties of our solar evolutionary models, we carry out seismic inversions of the squared adiabatic sound speed, $c^{2}=\frac{\Gamma_{1}P}{\rho}$, with $P$ the local pressure, $\rho$ the local density and $\Gamma_{1}=\frac{d \ln P}{d \ln \rho} \bigg|_{S}$ the first adiabatic exponent, of the entropy proxy, $S_{5/3}=\frac{P}{\rho^{5/3}}$ and the Ledoux discriminant, $\rm{A}= \frac{1}{\Gamma_{1}}\frac{d \ln \rm{P}}{d \ln \rm{r}}-\frac{d \ln \rho}{d \ln \rm{r}}$. The combined analysis of these helioseismic inversions allows us to have a clear view of the properties of solar models, as was carried out in \citep{Buldgen2019} and allows to investigate deeper the individual contributions of some key elements of solar models. We used the SOLA inversion technique \citep{Pijpers}, following the approach of \citet{RabelloParam} to calibrate the trade-off parameters. 

\subsection{Sound speed inversions}

\begin{figure}
	\centering
		\includegraphics[width=9cm]{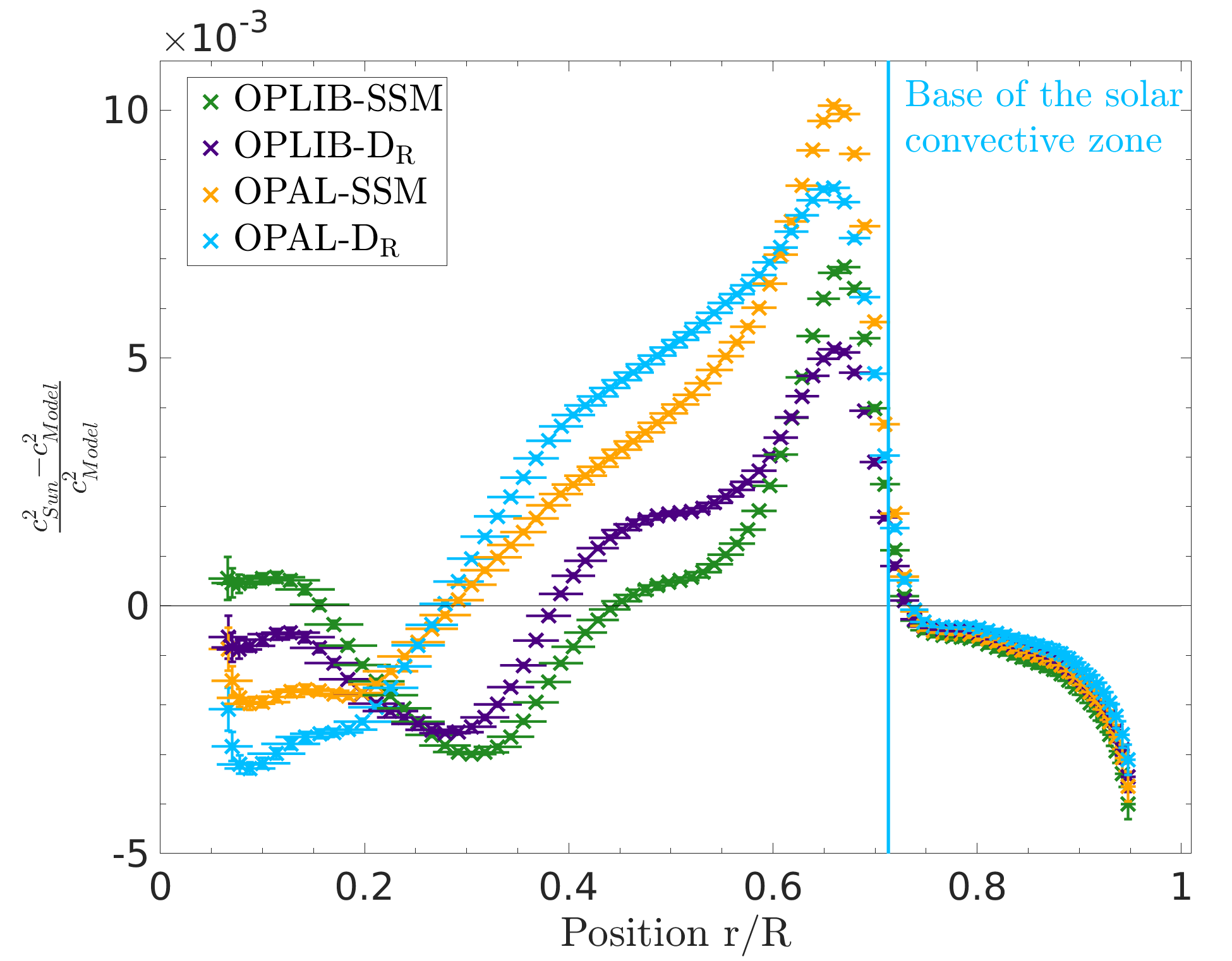}
	\caption{Relative squared adiabatic sound speed differences between the Sun and models using the OPAL and OPLIB opacities, either within the standard solar model framework or including macroscopic mixing of chemical elements.}
		\label{Fig:C2}
\end{figure}

The squared adiabatic sound speed inversions for all models (SSMs and models including macroscopic transport) is illustrated in Fig. \ref{Fig:C2}. As can be seen, varying the reference opacity tables for a given mixture has a significant impact, as was already illustrated in \citet{Buldgen2019} for the AGSS09 mixture. Looking only at the sound speed inversion, one could argue that the OPLIB SSM is superior to the OPAL SSM, especially in the upper radiative layers. A similar situation was found for GN93 abundances when using the OPLIB tables. However, given the issues regarding the neutrino fluxes mentioned above, these conclusions are incomplete and do not encompass the whole picture. Similarly, the inclusion of macroscopic transport leads to a slight improvement of the agreement at the BCZ, at the expense of increased discrepancies in the deeper layers, probably due to the reduction of the metallicity in the radiative zone. 

\begin{figure}
	\centering
		\includegraphics[width=9cm]{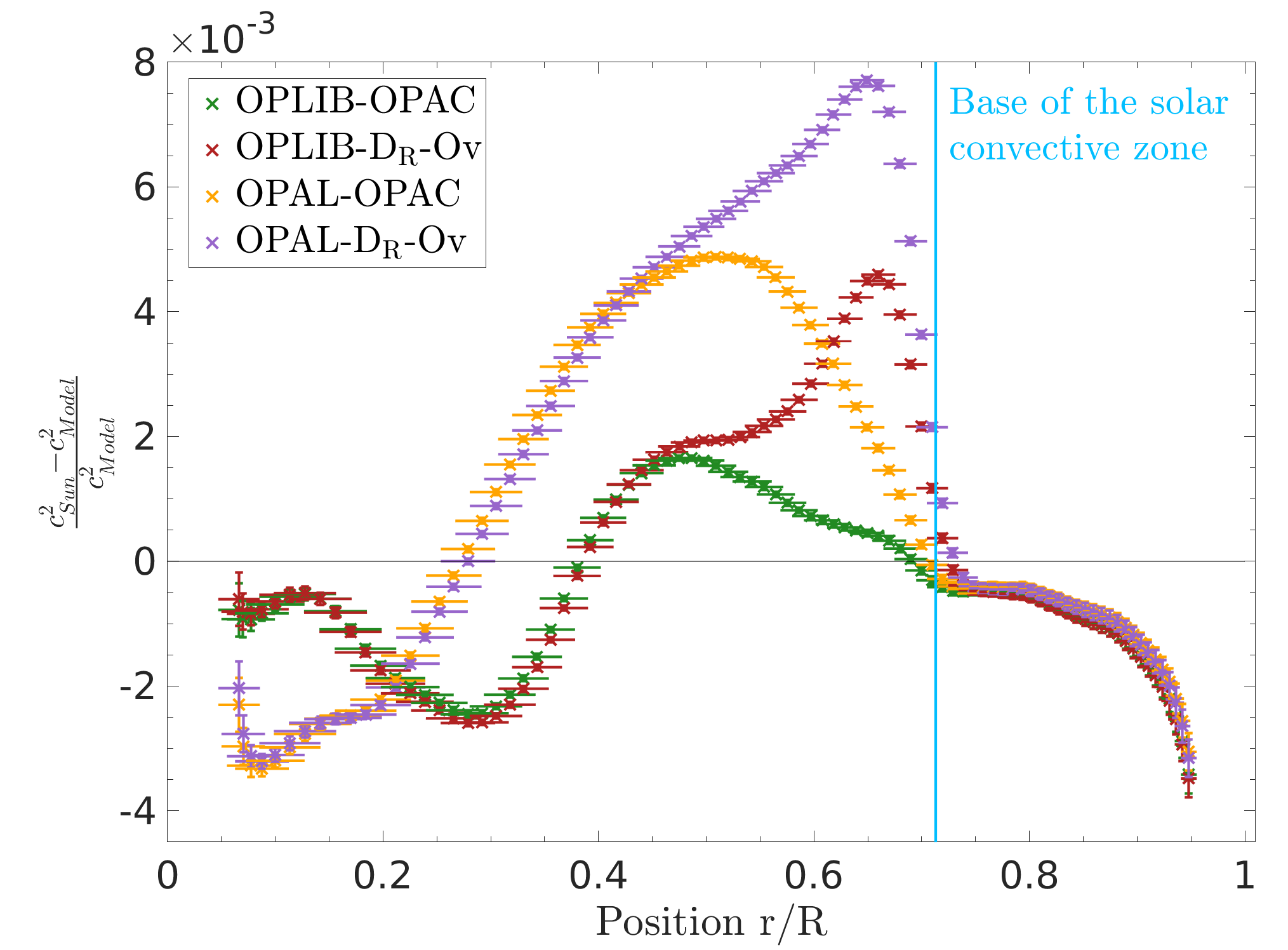}
	\caption{Relative squared adiabatic sound speed differences between the Sun and models using the OPAL and OPLIB opacities, including macroscopic mixing of chemical elements and either adiabatic overshooting or a localized opacity increase to replace the BCZ at the helioseismic value.}
		\label{Fig:C2OP}
\end{figure}

In Fig. \ref{Fig:C2OP}, we illustrate the impact of recovering the helioseismic value of $0.713$R$_{\odot}$ of the BCZ using either adiabatic overshooting or a localized increase of opacity. It appears that both approaches are not equivalent, and sound speed inversions would favour a localized opacity increase at the BCZ. Looking at other results in the literature \citep[e.g.][]{MonteiroOne,Rempel04,JCDOV,Zhang2019,Baraffe2022}, it appears that what is favoured is a significant change of temperature gradient on the radiative side. Further investigations would be required to see whether opacity modifications and effects of convective boundary mixing and thermalization of the convective elements can be distinguished. 

\subsection{Entropy proxy inversions}

\begin{figure}
	\centering
		\includegraphics[width=9cm]{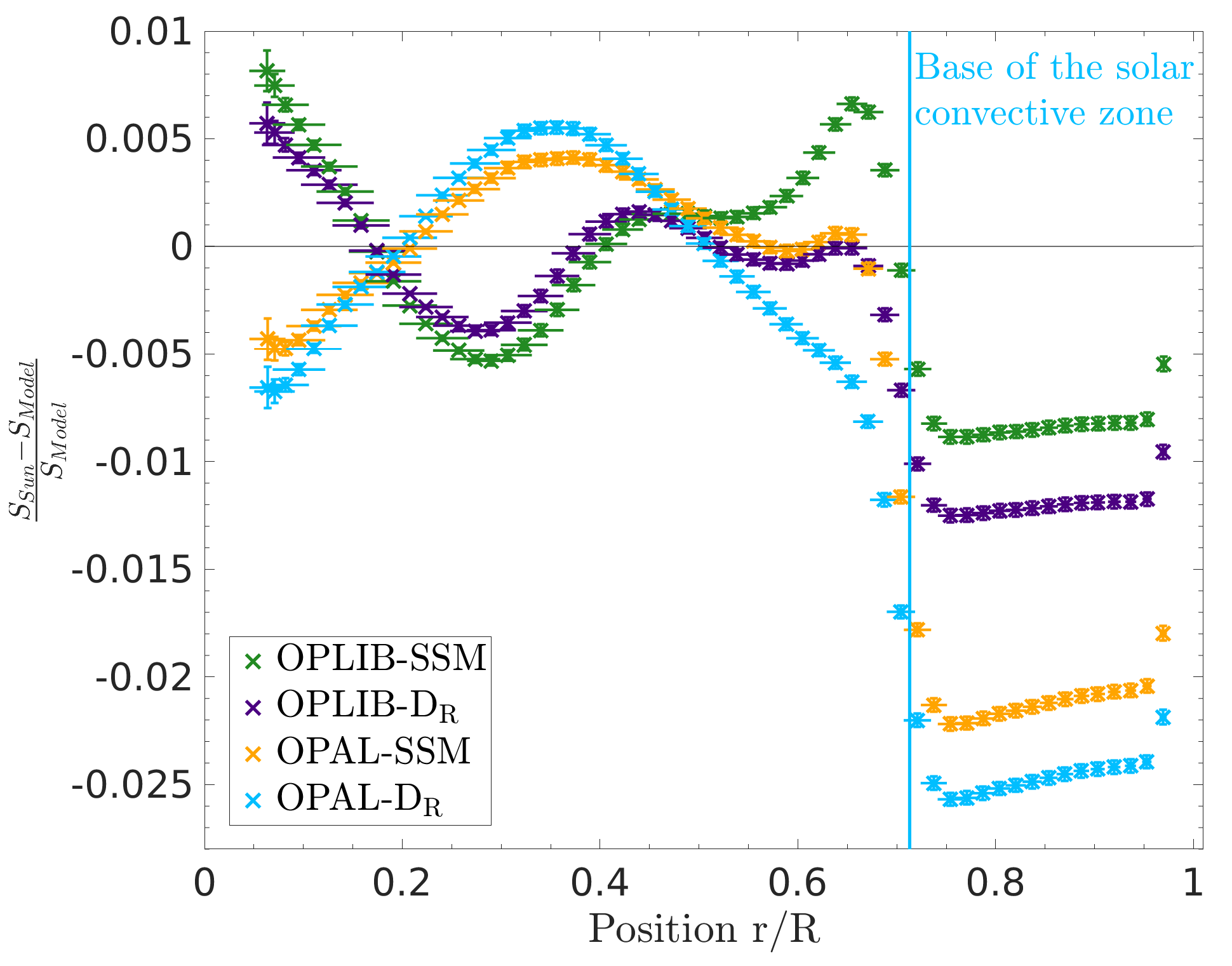}
	\caption{Relative entropy proxy differences between the Sun and models using the OPAL and OPLIB opacities, either within the standard solar model framework or including macroscopic mixing of chemical elements.}
		\label{Fig:S}
\end{figure}

The entropy proxy inversions for all SSMs and models including macroscopic mixing are illustrated in Fig. \ref{Fig:S}. Again, the best model seems to be the OPLIB SSM, which only shows small discrepancies throughout the radiative layers and a good agreement regarding the height of the entropy plateau in the CZ. This is in line with the conclusion of \citet{BuldgenS}, who observed a similar trend for both AGSS09 and GN93 abundances when comparing OPAL to OPLIB models. The inclusion of macroscopic transport significantly improves the agreement around $0.6$ R$_{\odot}$ for the OPLIB model, by essentially erasing the contribution of mean molecular weight gradients to this quantity. The situation is exactly the opposite for the OPAL model, as macroscopic transport leads to significiant discrepancies. Regarding the position of the entropy plateau in the CZ, macroscopic transport leads to a worsening of the agreement of about $0.003$, which is still significant at our precision level. It appears that none of the models, standard or not, are able to place the entropy plateau at the correct height. In the deeper radiative layers and the core, the changes remain quite small, overall similar to the sound speed variations.   

\begin{figure}
	\centering
		\includegraphics[width=9cm]{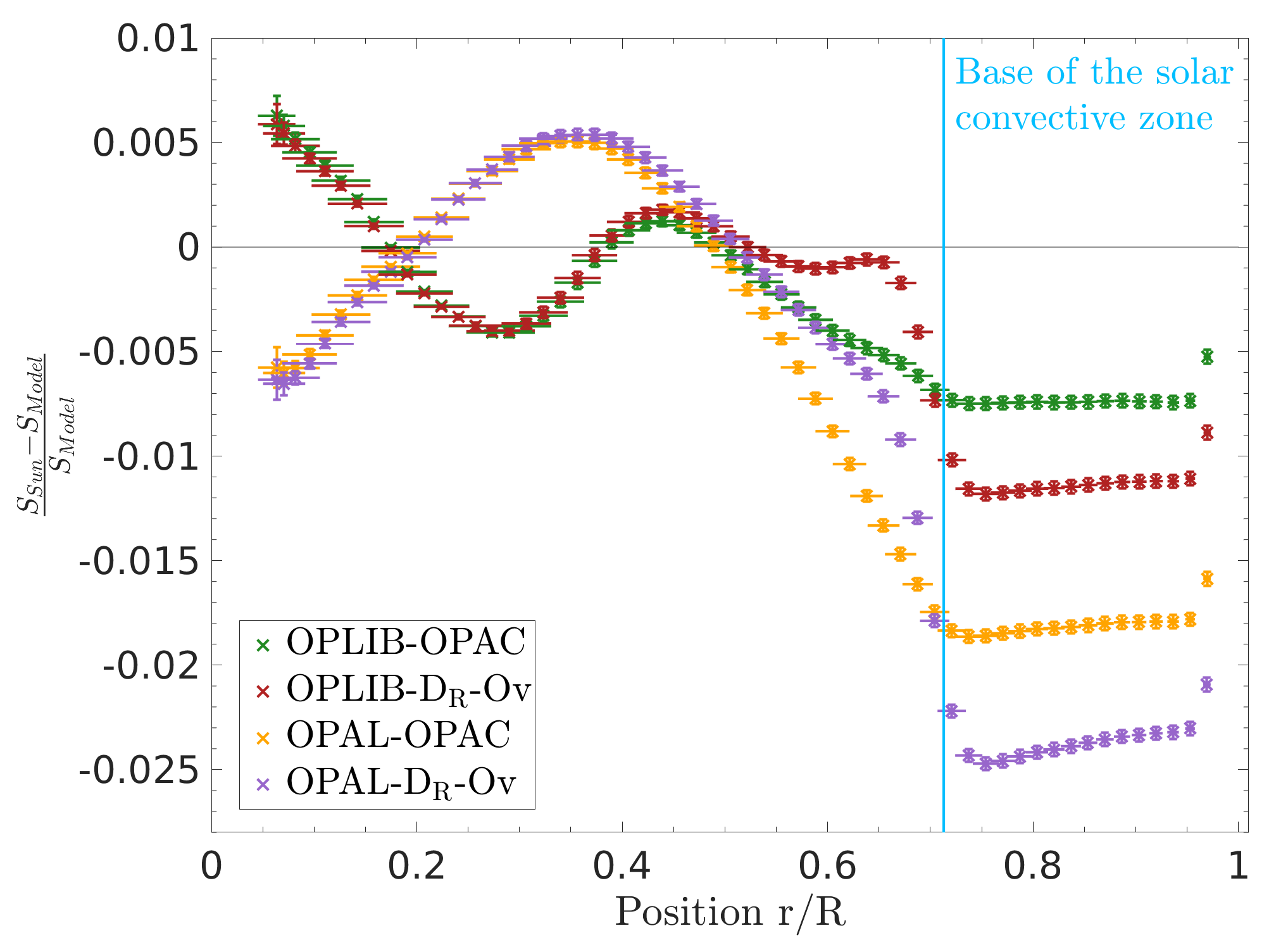}
	\caption{Relative entropy proxy differences between the Sun and models using the OPAL and OPLIB opacities, including macroscopic mixing of chemical elements and either adiabatic overshooting or a localized opacity increase to replace the BCZ at the helioseismic value.}
		\label{Fig:SOP}
\end{figure}

As for the sound speed inversion, the entropy proxy keeps a trace of how the position of the base of the convective envelope is replaced. It appears that including a localized opacity modification improves the height of the entropy plateau by about $0.005$, but worsen the agreement just below the BCZ, around $0.6$ R$_{\odot}$ (See Fig. \ref{Fig:SOP}). This conclusion is reached for both OPAL and OPLIB opacity tables and is likely due to the shape of the opacity modification. The inclusion of adiabatic overshooting however does not affect at all the height of the entropy plateau. In the case of AGSS09 abundances, \citet{Buldgen2019} found that adiabatic overshooting, if extended deep enough, could have a strong impact on the height of the plateau in one of their models, but the sound speed inversion results were then significantly worse. This implies that the temperature gradient in the radiative layers is poorly described with the current approach but that some degree of steepening is required to place the plateau in the CZ at the correct height. This contradicts the sound speed profile inversions who would strictly favour a model using the opacity modification of Eq. \ref{eq:OpalParam}.

From the entropy proxy inversions, the situation appears far more complex, despite the revision of the abundances by MB22 that improves the agreement from the point of view of the sound speed profile. 

\subsection{Ledoux discriminant inversions}

\begin{figure}
	\centering
		\includegraphics[width=9cm]{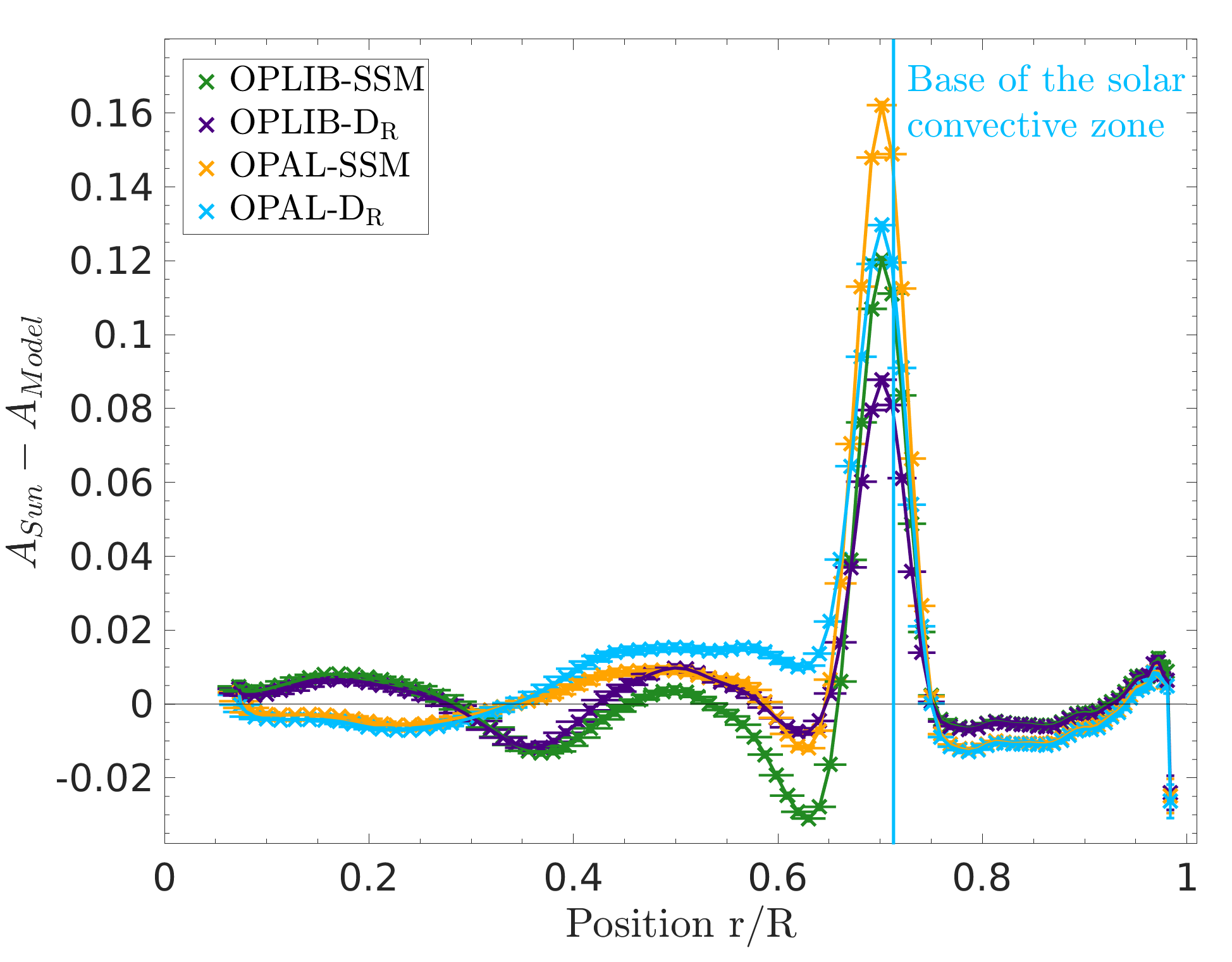}
	\caption{Ledoux discriminant differences between the Sun and models using the OPAL and OPLIB opacities, either within the standard solar model framework or including macroscopic mixing of chemical elements.}
		\label{Fig:A}
\end{figure}

The last inversion we investigate is that of the Ledoux discriminant profile, which amplifies the discrepancies at the BCZ. From Fig. \ref{Fig:A}, we see that the OPLIB SSM is only superior to the OPAL one close to the BCZ. In a similar fashion to what was observed for the GN93 abundances in \citet{BuldgenA}, there seems to be a region, around $0.6$ R$_{\odot}$, where the temperature gradient is too steep in the OPLIB model. This can be due to a too high opacity in these layers, as a result of the higher oxygen and iron abundance. Indeed, this discrepancy is located very close to the peak in metallicity that is observed in the SSMs due to the competing effects of pressure and thermal diffusion, that will in turn induce a higher opacity as the metals are the most dominant contributors at these temperatures \citep{BlancardOpacDetail}. The inclusion of macroscopic mixing erases this peak in metallicity and thus leads to a much less steep temperature gradient. For the OPLIB model, this improves significantly the agreement around $0.6$ R$_{\odot}$, but the improvement is much smaller for the OPAL model. The sharp variations at the base of the convective zone are slightly reduced by the presence of macroscopic mixing, indicating that a less steep chemical composition gradient is favoured, in agreement with previous studies \citep[e.g.][]{Brun2002,Takata2003,Baturin2015,JCD18}.  

\begin{figure}
	\centering
		\includegraphics[width=9cm]{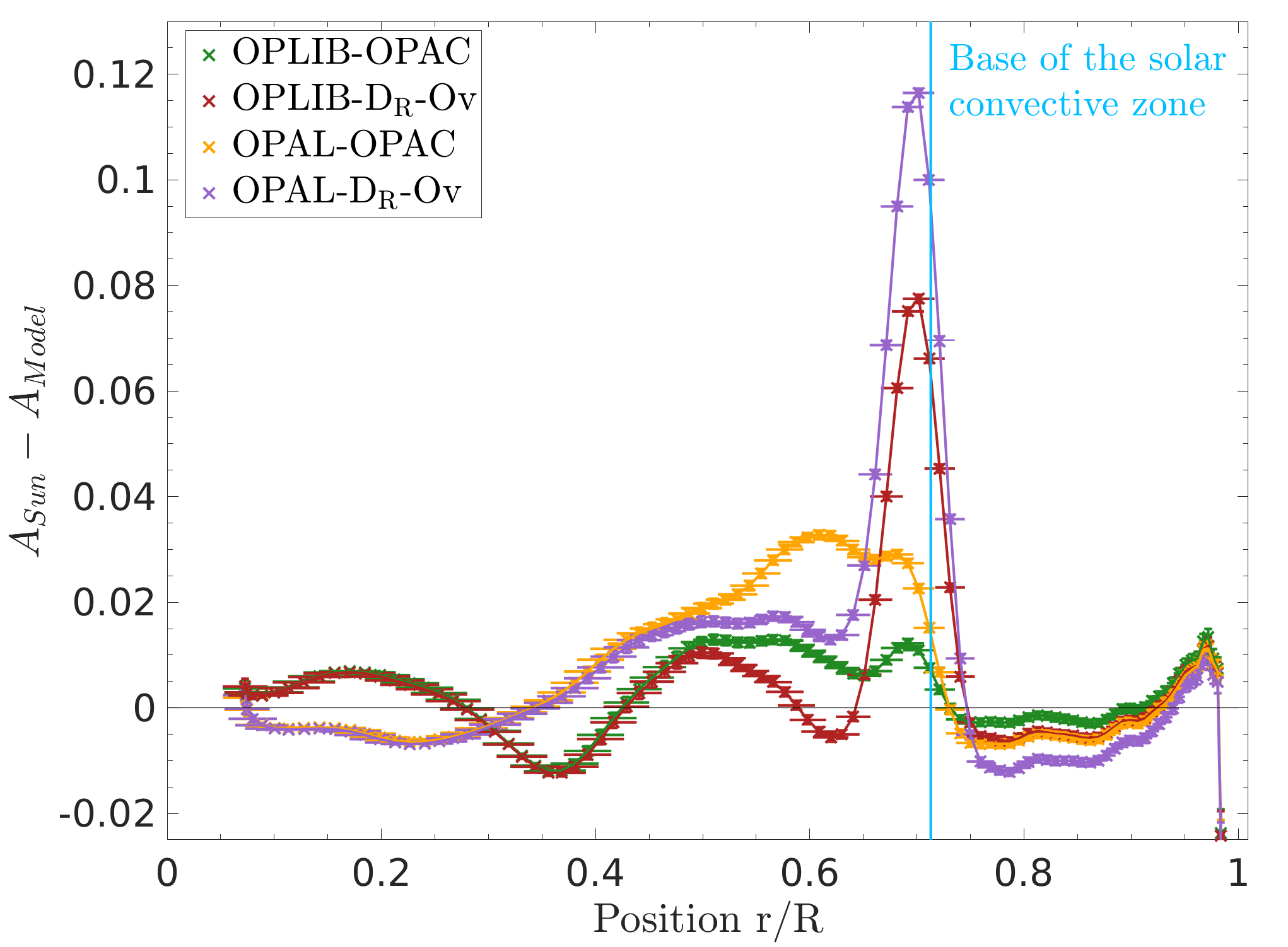}
	\caption{Ledoux discriminant differences between the Sun and models using the OPAL and OPLIB opacities, including macroscopic mixing of chemical elements and either adiabatic overshooting or a localized opacity increase to replace the BCZ at the helioseismic value.}
		\label{Fig:AOP}
\end{figure}

The effects of including adiabatic overshooting and a localized increase in opacity are illustrated in Fig. \ref{Fig:AOP}. Again both effects can be easily distinguished, with in both cases the opacity increase allowing to reduce efficiently the sharp peak at the BCZ. The OPLIB model with opacity increase is again favoured, while the OPAL model with opacity increase shows a quite extended deviation in the bulk of the radiative zone. This is due to the exact shape of the opacity profile in the model that extends to slightly higher temperatures in this case. Indeed, the amplitude of the opacity modification was increased from 8.5$\%$ to 12$\%$ between the OPLIB and the OPAL model without changing the width of the Gaussian function. Therefore the amplitude remains slightly larger at higher temperatures and explains the deviations, as illustrated in Fig. \ref{Fig:Kappa}. This demonstrates both that opacity modifications should remain very localized in these models and that the Ledoux discriminant inversion is extremely efficient at constraining the temperature gradient in the upper solar radiative layers. 

\subsection{Frequency separation ratios}\label{Sec:Ratios}

The frequency separation ratios defined by \citet{Roxburgh2003} are classical constraints in helioseismology. They have been used in numerous discussions related to the solar abundances \citep[e.g.][]{Basu2007,Chaplin2007} and the physics of solar models \citep{BuldgenS,Salmon2021}. They serve as a direct test of the sound speed gradient in the deep solar layers \citep{Roxburgh2003}, as they can be shown, using asymptotic developments \citep{Shibahashi1979,Tassoul1980} that they satisfy the following relation\\
\begin{align}
r_{n,\ell}\approx \frac{-(4 \ell + 6)}{4\pi^{2}\nu_{n,\ell}}\int_{0}^{R}\frac{dc}{dr}\frac{dr}{r}. 
\end{align}
with $c$ the adiabatic sound speed defined above, $\ell$ the degree of the mode and $R$ the solar radius. This constraint is regularly used for the asteroseismic modelling of solar-like oscillators but also as a straightforward test of solar models. \\

Therefore, we computed the frequency separation ratios for all the models in this study and compare them to a SSM using the GN93 abundances, FreeEOS and OPAL opacities, which would be the reference of the 90s to reproduce. To better illustrate the level of agreement we compare the following quantity
\begin{align}
\varepsilon_{n,\ell}=\frac{r^{Obs}_{n,\ell}-r_{n,\ell}^{\rm{Mod}}}{\sigma_{r_{n,\ell}}}
\end{align} 
with $\sigma_{r_{n,\ell}}$ the uncertainty on the observed frequency separation ratios. This quantity has the advantage of directly showing how significant the differences are. However, it should be kept in mind that the very high precision of the solar data implies that almost no models reach a $1\sigma$ level of agreement for all frequency separation ratios. 

\begin{figure*}
	\centering
		\includegraphics[width=17cm]{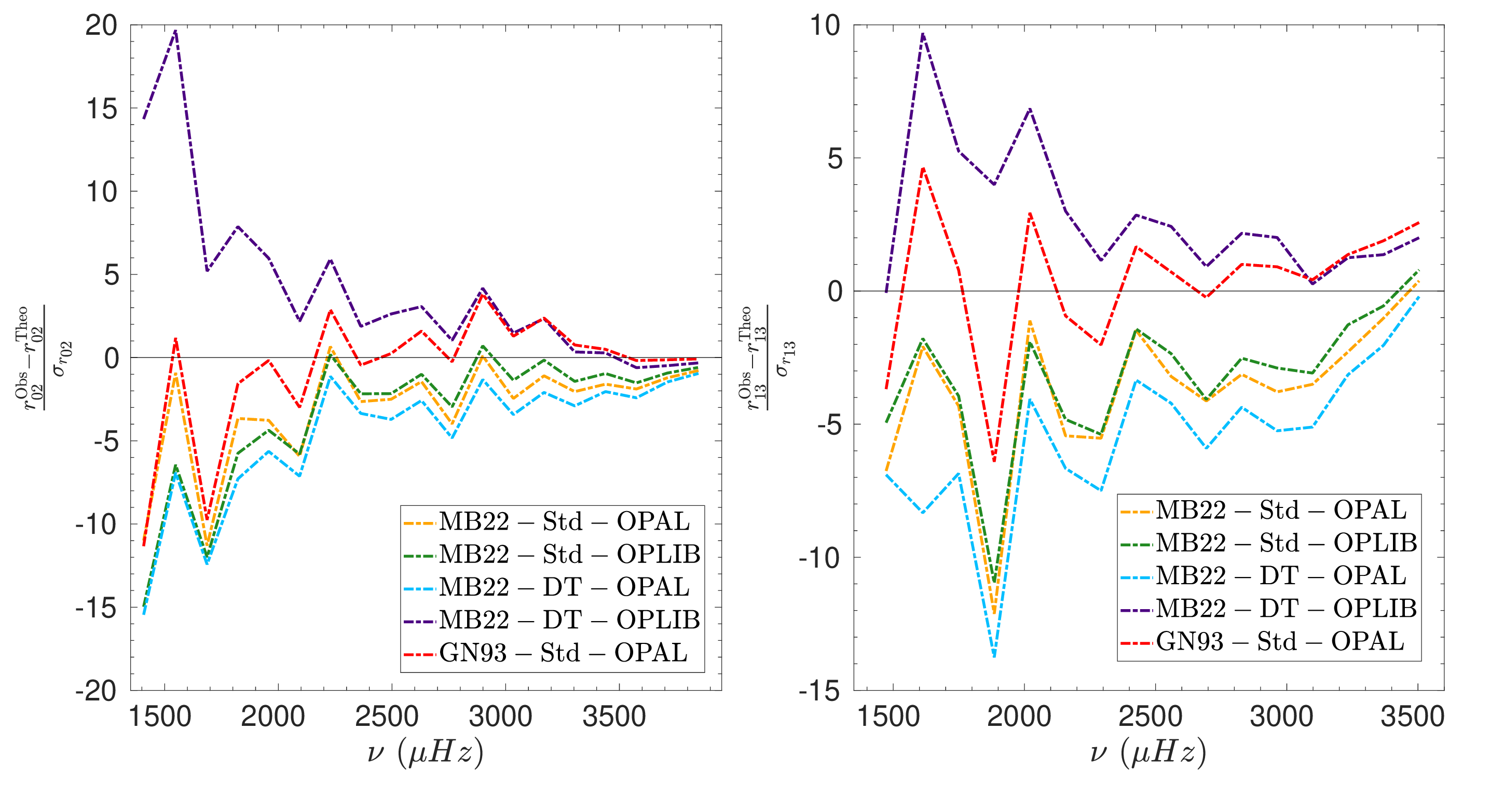}
	\caption{Frequency separation ratios as a function of frequency of standard solar models and models including macroscopic mixing described in Table \ref{tabModels}, compared to BiSON low degree data. A standard solar model using the GN93 abundances is also shown in comparison.}
		\label{Fig:ratiosStd}
\end{figure*}

\begin{figure*}
	\centering
		\includegraphics[width=17cm]{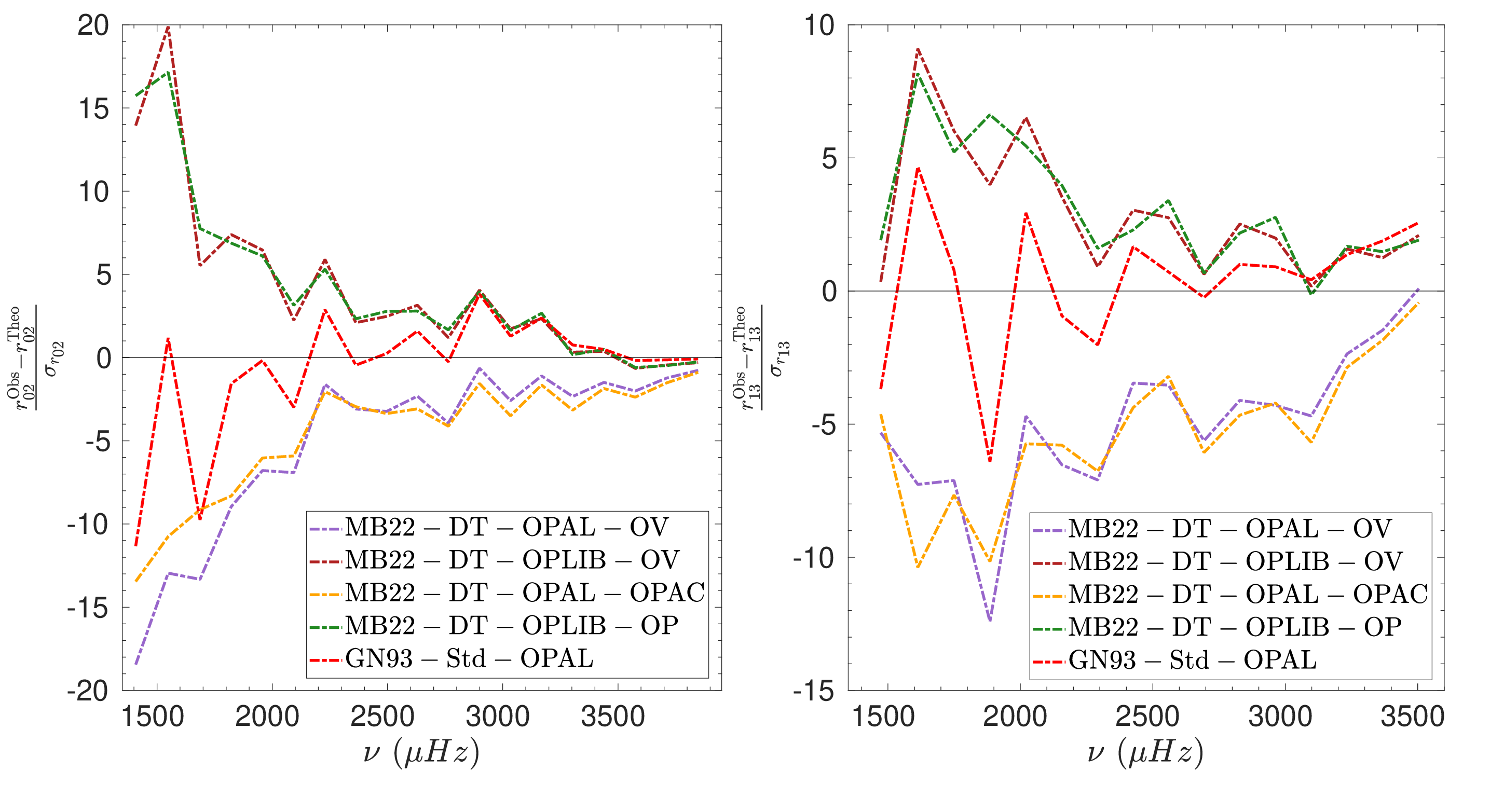}
	\caption{Frequency separation ratios as a function of frequency of models including macroscopic mixing and an localized opacity increase or convective overshooting at the BCZ described in Table \ref{tabModels}, compared to BiSON low degree data. A standard solar model using the GN93 abundances is also shown in comparison.}
		\label{Fig:ratiosOVOP}
\end{figure*}

These results are illustrated in Fig. \ref{Fig:ratiosStd} for the standard solar models and models with the revised turbulent formalism and Fig. \ref{Fig:ratiosOVOP} for the models including both macroscopic transport and overshooting or an opacity modification. From both Figs. \ref{Fig:ratiosStd} and \ref{Fig:ratiosOVOP}, none of the models performs quite well, whathever the opacity table used for them. If we compare the MB22 models to the GN93 model, we can clearly see that their level of agreement is significantly lower, particularly at low and intermediate frequencies, for both OPLIB and OPAL opacities, even after replacing the BCZ position using overshooting or an opacity increase.

This situation is in clear contrast with the agreement found for higher metallicity models of the 90s \citep[See e.g.][]{Chaplin2007,Basu2007,Serenelli2009,BuldgenS}. Due to the similarities between the OP and the OPAL opacities, the same level of agreement is found if the experiment is repeated with an OP model. Further investigations regarding the exact layers to which the frequency separation ratios, and their slope, are sentitive to could be perhaps help pinpoint the exact origins of the observed deviations. Indeed, AGSS09 models using the OPLIB opacities are found to provide a good agreement while they are clearly in disagreement regarding other key constraints such as neutrino fluxes and helium abundance in the convective zone \citep[see][and the associated discussion]{BuldgenS,Salmon2021}. 

\section{Discussion}\label{Sec:Disc}

The immediate takeaway from our detailed analysis of solar models with revised MB22 abundances is that the agreement found for SSMs strongly depends on the radiative opacities used in their computation. While models using OPAL tables show an excellent agreement with helioseismic and neutrino data, models using OPLIB tables show significant discrepancies in neutrino fluxes while showing overall better agreement with helioseismic constraints, with the exception of the helium mass fraction in the CZ. This issue regarding the OPLIB tables was already discussed in \citet{Buldgen2019} and \citet{Salmon2021} and detailed comparisons are required to determine the origin of the differences between OPAL, OP and OPLIB. Similarly, the good performance of the OPAL opacities are not surprising given the small differences between OPAL and OP. Therefore, even for SSMs, a detailed analysis of the models leads to conclude that despite the improvements due to the increase of the solar abundances, key issues remain linked to the current state of solar modelling. In this context, the existing degeneracies in classical helioseismic inferences cannot be used to validate abundance determinations. The conclusions we draw from the OPAL models can thus be applied to the OP models, given the similarities between the two opacity tables. A striking issue is also found for the frequency separation ratios, which only provide a moderate agreement with solar models using the MB22 abundances, far from what was achieved with GS98 or GN93 models. In that respect the MB22 abundances are not exactly equivalent to the GS98 abundances, further investigations are required to examine exactly from where these discrepancies arise.

The second main takeaway is that the situation drastically changes once macroscopic mixing is taken into account to reproduce the lithium depletion at the solar surface. This was already discussed in \citet{Buldgen2023} and is generalized here to models including the OPAL and OPLIB opacities. The significant decrease in neutrino fluxes observed for models including macroscopic mixing is due the drastic change in the calibration results once one attempts to reproduce the lithium observations. As mentioned in \citet{Buldgen2023}, a reliable beryllium determination would be required to further constrain the efficiency of macroscopic mixing at the BCZ. The impact of planetary formation could mitigate the issue \citep{Kunitomo2022}, but modifications to other key physical ingredients such as opacity at higher temperatures and electronic screening \citep{Mussack2011A,Mussack2011B} cannot be excluded.  

The third main takeaway is linked to the effect of macroscopic mixing on the position of the BCZ and its effect on key indicators of thermal gradients such as the entropy plateau in the CZ and the Ledoux discriminant. We confirm that helioseismic data strongly favours significant modifications of the thermal gradients in the radiative zone even with revised abundances. Replacing the BCZ to the helioseismic value of $0.713$R$_{\odot}$ using adiabatic overshooting does not significantly improve the agreement of the models with helioseismic inferences, whereas a localized increase of opacity decreases the observed discrepancies. Whether the changes in the temperature gradients can be due to the thermalization of the convective elements in the radiative zone remains to be explored using physically motivated prescriptions \citep{Baraffe2022}. However, a key difference between an opacity increase and effects linked to convective overshooting resides in the mixing of chemical elements and the potential impact on lithium and beryllium depletion. On the other hand, the opacity modification implemented here is not totally realistic, as an actual revision of opacities might lead to significant changes in opacity at higher temperatures, as was seen for the OPAS and OPLIB tables and can be expected from new computations \citep[e.g.][]{Nahar2016,Zhao2018,Pain2020,Pradhan2018,Zhao2018,Pradhan2023}. 

Overall, the situation appears to be quite complex and still requires extensive investigations. While uncertainties between the various opacity tables are worrying, the dispersion of $0.6$ dex inferred by MB22 for iron (Table A.1, Appendix A), not seen by \citet{Asplund2021}, significantly worsen the situation as this element is, with oxygen, the first contributor to opacity at the BCZ and remains highly significant throughout the solar radiative zone. Given the uncertainties on iron opacity \citep{Bailey} and the impact of new physical processes in the computations \citep{Pradhan2023}, a detailed discussion on these discrepancies is required before a definitive conclusion regarding solar models can be reached. Further analyses using linear solar models \citep{Villante2010}, seismic models \citep{Buldgen2020} or extended calibration procedures \citep{Ayukov2017,Kunitomo2021} might be informative, but these will require to be combined with theoretical inputs to lift the degeneracies that appear once non-standard solar models are used in the solar calibration procedure. Such degeneracies are not present in the SSM calibration that used a simplified physical picture.

\section{Conclusion}\label{Sec:Conc}

In this study, we have analyzed in details solar models computed with the abundances proposed by MB22. We investigated the effects of changing the opacity tables within the SSM framework, including macroscopic mixing at the BCZ mimicking the effects of rotating models \citep{Eggenberger2022}, then adding the adiabatic overshooting or a localized opacity increase to compensate for the effects of macroscopic mixing on the position of the BCZ. A complete picture of the situation is drawn by looking at the global properties and neutrino fluxes of the models in Sect. \ref{Sec:Models} as well as combined helioseismic inversions in Sect. \ref{Sec:Inversions}. The results are discussed in Sect. \ref{Sec:Disc}, where three main points of discussion are outlined. 

We conclude that the proposed revision of the solar abundances by MB22 does not change the need for future improvements of solar models. While MB22 consider that the remaining "discrepancies" can be solved using investigations following \citet{Buldgen2019}, we show in this study that this is not the case. On the contrary, it appears that the good agreement regarding sound speed, neutrino fluxes and global parameters found by MB22 for their SSMs is due to a favourable combination of physical ingredients of their models. In a similar fashion to a tightrope walker, a small push in a given direction worsens the situation for the SSMs. For example, changing the radiative opacities or including macroscopic transport, leads to an overall worsening of the situation regarding helium, neutrino fluxes or BCZ position while significantly changing the results of helioseismic inferences, not always for the better. In addition, choices regarding spectral lines, datasets and microphysical ingredients in the spectroscopic analysis by MB22 need to be discussed, as well as the extreme spread in iron abundance they find that would drastically change the properties of solar models. 

A clear difference between low and high-Z solar models is that further improvements of the physics of the models, such  as an opacity increase motivated by recent works, or the inclusion of light-element depletion, tend to reduce some of the discrepancies in low-Z models, while they increase them in high-Z models. This is to be put in perspective in the context of recent solar envelope metallicity determinations \citep{Vorontsov13,BuldgenZ, Buldgen2023Z} which tend to be consistent with AAG21 spectrosopic values, leaving the differences with MB22 to be explained. 

As outlined in \citet{Buldgen2019}, renewed detailed analyses of solar models are required to determine the importance of numerical uncertainties in the comparisons of solar models with the highly precise constraints available for the Sun. In parallel, experimental efforts for more precise determinations of CNO neutrino fluxes as well as experimental and theoretical opacity values in solar conditions remain key factors to constrain the deep radiative interior of solar models. Regarding the solar convective layers and the BCZ interface, our work shows that combining helioseismic inversions to light element depletion might provide a data-driven analysis of both the chemical and thermal properties of the BCZ. Further improvements to the resolution of the inversion techniques, by using non-linear RLS methods \citep{Corbard1999} might however be required to get a full picture. 

\section*{Acknowledgements}

We thank the referee for their thorough reading of the manuscript. G.B. is funded by the SNF AMBIZIONE grant No 185805 (Seismic inversions and modelling of transport processes in stars). A.M.A. gratefully acknowledges support from the Swedish Research Council (VR 2020-03940). P.E. received funding from the European Research Council (ERC) under the European Union's Horizon 2020 research and innovation programme (grant agreement No 833925, project STAREX). We acknowledge support by the ISSI team ``Probing the core of the Sun and the stars'' (ID 423) led by Thierry Appourchaux. 

\bibliography{biblioarticleMaggComb}

\end{document}